%% file: EFT_E4.tex
\documentclass[12pt]{article}
\usepackage[top=0.7in,bottom=1in,left=1.1in,right=1.1in,includehead]{geometry}
\usepackage{epsfig,amsmath,amsfonts,verbatim}
\usepackage{lmodern}
\usepackage[T1]{fontenc}
\usepackage{mathtext,marvosym,textcomp}
\usepackage{indentfirst}
\usepackage{epsfig,amsmath,amsfonts}
\usepackage{mathtools}
\usepackage[usenames,dvipsnames,svgnames,table]{xcolor}
\usepackage{tikz}
\usetikzlibrary{arrows}
\usetikzlibrary{shapes.misc}

\tikzset{cross/.style={cross out, draw=black, minimum size=2*(#1-\pgflinewidth), inner sep=0pt, 
outer sep=0pt},
cross/.default={5pt}}
 \usepackage[ normalem]{ulem}
 \usepackage{tocloft}
 
 \usepackage[debug,pageanchor=false]{hyperref}
\hypersetup{colorlinks=true,linktocpage,breaklinks,
            urlcolor=blue,
            linkcolor=red,
            citecolor=blue
            }

\numberwithin{equation}{section}

\input{Defs.tex}

\begin{document}
\renewcommand{\contentsname}{}
\renewcommand{\refname}{\begin{center}References\end{center}}
\renewcommand{\abstractname}{\begin{center}\footnotesize{\bf Abstract}\end{center}} 
 \renewcommand{\cftdot}{}

\begin{titlepage}

\ph{preprint}

\vfill

\begin{center}
   \baselineskip=16pt
      {\Large \bf  Exceptional field theory: $SL(5)$.}
      \vskip 2cm
        Edvard T. Musaev\footnote{\tt edvard.musaev@aei.mpg.de}
          \vskip .6cm
                \begin{small}
                             {\it National Research University Higher School of Economics,\\
                              Faculty of Mathematics \\
                             7, st. Vavilova, 117312, Moscow, Russia} \\[0.5cm]
                           {\it  Max-Planck-Institut f\"ur 
                                Gravitationsphysik (Albert-Einstein-Institut),\\
                           Am M\"uhlenberg 1, 14476 Potsdam, Germany}
                \end{small}
\end{center}

\vfill 
\begin{center} 
\textbf{Abstract}
\end{center} 
\begin{quote}
In this work the exceptional field theory formulation of supergravity with SL(5) gauge group is considered. This group appears as a U-duality group of $D=7$ maximal supergravity. In the formalism presented the hidden global duality group is promoted into a gauge group of a theory in dimensions 7+number of extended directions. This work is a continuation of the series of works for $E_{8,7,6}, SO(5,5)$ and $SL(3)\times SL(2)$ duality groups.
\end{quote} 
\vfill
\setcounter{footnote}{0}
\end{titlepage}

\clearpage
\setcounter{page}{2}

\tableofcontents

\section{Introduction}

The construction of Extended Geometry appeared in a series of works 
\cite{Siegel:1993th,Hull:2007zu,Hull:2009zb,Berman:2010is,Berman:2011jh,Berman:2011pe} and many 
others has got a 
lot of attention recent years in particular in the context of string cosmology 
\cite{Hassler:2014mla,Blumenhagen:2015lta}, searches for non-geometric solutions 
\cite{Dibitetto:2012ia,Dibitetto:2012rk,Hassler:2013wsa,Jensen:2011jna,Andriot:2012an,
Andriot:2012wx,Andriot:2011uh}, 
generalized Scherk-Schwarz compactifications and embeddings of Type II solutions 
\cite{Grana:2012rr,Berman:2012uy,Musaev:2013rq,Blair:2013gqa,Baron:2014yua,Malek:2015hma,
Baguet:2015sma,
Baguet:2015xha} and many 
others. The basic idea of the model is to turn the hidden U-duality symmetries of 
compactified (half-)maximal supergravities into a manifest gauge symmetry of a full 
non-compactified theory. This is achieved by introducing new coordinates in addition to the 
existing 10 or 11. In the string or M-theory interpretation these new coordinates correspond to 
winding modes of strings and D- or M-branes (see e.g.\cite{Hull:2007zu}). Geometric meaning of 
the extended 
space was investigated in the works 
\cite{Coimbra:2011ky,Coimbra:2011nw,Coimbra:2012af,Coimbra:2014uxa,
Cederwall:2014opa}. Very important direction of research is bound to the question of finite 
coordinate transformations in the extended space. To the moment it is not clear how to integrate 
infinitesimal generalized diffeomorphisms to a large transformation, however there has been certain 
progress made \cite{Berman:2014jba,Cederwall:2014kxa,Rey:2015mba}. In the work \cite{Palmkvist:2015dea} attempts to construct the algebra without the need of section condition were made. Construction of superspace for 
DFT was considered in the work 
\cite{Bandos:2015cha}. In the series of works \cite{Berman:2014hna,Berman:2014jsa,Berkeley:2014nza} 
it was shown that the string and M-theoretical extended objects can be understood as wave-like or 
monopole-like solutions of Exceptional Field Theory.

Since the particular U-duality group $E_{n(n)}$ depends on the number of compact directions $n$ 
 the full set of coordinates of EFT naturally splits into the so-called ``external'' 
space-time coordinates denoted by $x^\m$ and ``internal'' extended coordinates denoted by $\XX^M$. 
Here $\m$ runs over all would-be non-compact directions, i.e. from 1 to $11-n$ (for the maximal 
case), while the capital Latin index labels the necessary representation of the U-duality group, that catches all the translational and winding modes of the M2 and M5 branes. 

Although, EFT is formulated as a non-compactified theory, one has always keep in mind an additional 
condition, that forces us to either drop all or a subset of the extended coordinates returning 
effectively to the conventional supergravity or its toroidal compactifications, or to perform a 
generalized Scherk-Schwarz reduction, leading to the (half)maximal gauged supergravity 
\cite{Grana:2012rr,Berman:2012uy,Musaev:2013rq,Baron:2014yua}.

This condition, named \emph{section condition}, appears in the algebra of generalized 
diffeomorphisms given by the so-called generalized Lie derivative \cite{Berman:2012vc}. These may be 
understood as 
local U(T)-duality transformations and the theory should be (co)invariant with respect to them. For 
more details the reader is referred to the reviews 
\cite{Aldazabal:2013sca,Berman:2013eva,Hohm:2013bwa}.

Special properties of generalized Lie derivatives allow to obtain the 
tensor hierarchy of gauged supergravity as a natural consequence of the algebra and the section 
condition. It was proposed in \cite{Hohm:2013jma,Hohm:2013pua} to let all the gauge parameters to 
depend on the whole set of 
coordinates $(x^\m,\XX^M)$, and use the vector field $A_\m^M$ of the corresponding supergravity to 
introduce a derivative covariant with respect to generalized Lie transformations. Due to failure of 
the Jacobi identity for these transformations one is required to deform the 2-form field strength 
in the spirit of tensor hierarchy introducing a 2-form potential. Bianchi identities uniquely 
determine the corresponding field strength and the sequence can be continued.

Based on these ideas is the Exceptional Field Theory formalism 
presented in the series of papers \cite{ 
Hohm:2014fxa,Hohm:2013uia,Hohm:2013vpa,Abzalov:2015ega,Hohm:2015xna} for 
the groups E$_{8,7,6}$, SO(5,5) and SL(3)$\times$SL(2), that correspond to $D=3,4,5,6$ and 8 
maximal supergravities. The supersymmetric construction that includes fermions is given in
 \cite{Godazgar:2014nqa,Musaev:2014lna,Musaev:2015pla}. In 
this paper we aim at the construction for $D=7$ and the group SL(5). To the moment there has been 
large progress in investigation of the SL(5) extended geometry 
\cite{Berman:2011kg,Berkeley:2015mmc}. To be noted is the work \cite{Park:2014una} that considers 
the internal sector, i.e. the so-called scalar potential, for SL(N) group for any $N$. In the current paper we present the full EFT construction, including the gauge kinetic and topological sector.

This paper starts with the section \ref{Content}, where the bosonic field content of the maximal supergravity in D=7 is discussed. In section \ref{EG_hier} we briefly review the construction of exceptional field theory and tensor 
hierarchy to set up our notation for further reference. In section \ref{EFT} we construct a duality 
covariant kinetic and topological Lagrangian and check its invariance under external 
diffeomorphisms. We comment there on the feature of EFT when all the numerical prefactors in the 
Lagrangian are fixed already at the bosonic level. Embedding of the 11-dimensional and Type II 
supergravities into the formalism is considered in  section \ref{Embed}. Finally, in the 
Appendix we present the notations used, conventions on the SL(5) algebra, provide some useful 
identities and explicitly check gauge invariance of the topological Lagrangian.

\section{Supergravity in $D=7$}
\label{Content}

Maximal ungauged supergravity in 7 dimensions with the global duality group $\mathrm{SL}(5)$ was 
constructed in~\cite{Sezgin:1982gi}. The field content after the $7+4$ split is given 
by
\begin{equation}
\label{split}
\left\{ g_{\m\n}, A_{\m\,a}, \ff_{ab}, C_{\m\n\r}, B_{\m\n\,a}, A_{\m\,ab}, \ff_{abc}  \right\},
\end{equation}
where the internal indices $a, b$ run from 1 to 4. We have a total of ten 
1-form fields, $A_{\m\,a}$ and $A_{\m\,ab}$, which transform in the $\overline{\mathbf{10}}$ 
representation of the duality group $\mathrm{SL}(5)$ (see \cite{Cremmer:1997ct} for more details on how the supergravity fields are organized into irreps of U-duality group). We will denote this representation as 
$A_\m^M$, $M=1,\ldots,10$, where $M$ may also be thought of as a pair of $\mathrm{SL}(5)$ 
fundamental indices, $A_\m^M = \frac12 A_\m^{[ij]}$, $i,j = 1,\ldots,5$. More details on how we treat the doubled indices can be found in the Appendix \ref{Notations}.

The duality relation between 3-forms and 2-forms in seven space-time dimensions allows to collect the four 
fields $B_{\m\n\,a}$ and the 3-form $C_{\m\n\r}$ together, resulting in five 2-form fields $B_{\m\n\,i}$ 
transforming in the fundamental representation of $\mathrm{SL}(5)$.

There are now 14 scalar fields, $\ff_{ab}$ and $\ff_{abc}$, whose dynamics may be formulated in 
terms of the matrix $V_i^{\a\b}$ parameterizing the coset $\mathrm{SL}(5)/\mathrm{SO}(5)$. The small Greek indices $\a,\b = 1,\ldots, 4$ label the fundamental representation of $\mathrm{USp}(4) \simeq 
\mathrm{Spin}(5)$. We require the scalar matrix $V_i^{\a\b}$ to be antisymmetric in $\a,\b$ and 
traceless with respect to the $\mathrm{USp}(4)$ invariant tensor $\W_{\a\b}$, 
$V_i^{\a\b}\,\W_{\a\b}=0$~\cite{Samtleben:2005bp}. These constraints cut the number of degrees of 
freedom of $V_i^{\a\b}$ down to 25. Imposing additionally that $\det V = 1$ we constrain $V$ to 
have 
the right number of degrees of freedom and to be an $\mathrm{SL}(5)$ element. To respect the  tracelessness condition,the inverse of $V_i^{\a\b}$ is defined by the following identities
\begin{equation}
V_i^{\a\b}\, V_{\a\b}^j = \d_i^j,\qquad V_i^{\a\b}\, V_{\g\d}^i = \d_{\g\d}^{\a\b} - 
\frac14\,\W^{\a\b} \W_{\g\d}.
\end{equation}
One defines the SL(5) generalized metric as  $m_{ij} = V_i^{\a\b} V_{j\,\a\b}$, but sometimes it will be 
convenient to use the generalized metric in the $\bf 10$ of SL(5), defined as
\begin{equation}
\label{M_SL}
\mc{M}_{MN} \Longrightarrow \mc{M}_{ij,kl} =m_{ik} m_{jl} - m_{il} m_{jk}.
\end{equation}

As in the SO(5,5) case, duals for the 2-forms must be introduced as independent fields, in order to 
facilitate the description of different possible gaugings. Thus we introduce a set of 3-form 
fields $C_{\m\n\r}{}^i$, transforming in the $\overline{\mathbf{5}}$ of $\mathrm{SL}(5)$. These are related by a duality condition that will arise as an equation of motion. It will 
be convenient to redefine the 2- and the 3-form fields with the indices labeling the 
$\overline{\mathbf{10}}$ of $\mathrm{SL}(5)$, $B_{\m\n}{}^{KL}$ and $C_{\m\n\r}{}^{N,KL}$:
\begin{equation}
 \begin{aligned}
   B_{\m\n i}&=2\e_{iklmn}B_{\m\n}{}^{klmn},\\
   C_{\m\n\r}{}^m&=-6\e_{nklrs}C_{\m\n\r}{}^{mn,klrs}.
 \end{aligned}
\end{equation}

\section{Tensor hierarchy and Bianchi identities}
\label{EG_hier}

Let us now briefly review the EFT construction and setup our conventions. Transformation of 
covariant objects in Exceptional Field Theory is defined by the usual rule
\begin{equation}
\label{Lie}
\begin{split}
&\d_\L V^M=(\mc{L}_\S V)^M=(L_\L V)^M+Y^{MN}_{KL}\dt_N\L^K V^L\equiv[\L,V]_D^M, \\
\end{split}
\end{equation}
where $[\, ,]_D$ denotes the Dorfman bracket. For the SL(5) U-duality group the $Y$-tensor is given by
\begin{equation}
\label{Y}
Y^{MN}_{PQ}= \e^{m MN}\e_{m PQ},
\end{equation}
where $\e^{mMN}$ is an SL(5) invariant tensor whose components are given by the alternating 
symbol $\e^{mklpq}$. Note, that each large Latin letter parametrizes $\bf 10$ representation 
of SL(5) and through the paper is always equivalent to a pair of small Latin indices 
parameterizing $\bf 5$ of SL(5) (see Appendix \ref{Notations}).

Since the $Y$-tensor is related to the projector on the adjoint it is straightforward to write  
the generalized Lie derivative of a generalized vector in the following form
\begin{equation}
\d_\L V^M=(\mc{L}_\L V)^M=\L^N\dt_N V^M- 3{} \mathbb{P}^M{}_L{}^N{}_K\dt_N \L^K V^L + \fr15 (\dt_K 
\L^K)V^M.
\end{equation}
Here the last term plays the role of a weight term, that could be in principle added to any transformation. However, here it directly follows from the algebra. 
Rewriting the projector $\PP$ explicitly as in the Appendix \ref{algebra} one obtains the 
following transformation of a field in the fundamental $\bf 5$ representation
\begin{equation}
 (\mc{L}_\L U)^m=\L^N\dt_N U^m- \fr14 (t^m_n)^{kl}_{pq}\dt_{kl} \L^{pq} U^n + \fr{1}{10} (\dt_K 
\L^K)U^m.
\end{equation}
The weight $\l(U^m)=1/10$ is a half of that for $V^{mn}$ as it should be since one may 
always introduce a tensor $U^{[m}U^{n]}$ in $\bf 10$ that has the weight $2\l(U^m)$. In what 
follows it will prove useful to have transformation rules for the tensor $B_m=\e_{mKL}B^{KL}$ 
obtained by contraction of a 2-rank generalized tensor of weight $\l(B^{KL})=2/5$ with the 
epsilon-tensor. The resulting generalized tensor belongs to the $\bar{\bf 5}$ representation and 
transforms as
\begin{equation}
 (\mc{L}_\L B)_m=\L^N\dt_N B_m+ \fr14 (t^n{}_m)^{kl}_{pq}\dt_{kl} \L^{pq} B_n + \fr{2}{5} (\dt_K 
\L^K)B_m.
\end{equation}
As expected, this differs from the above transformation for a tensor in $\bf 5$ only by the weight term.

It is important to mention the section condition $Y^{MN}_{KL}\dt_M\otimes \dt_N=0$, that for the 
case of SL(5) U-duality group can 
be written as
\begin{equation}
 \e^{imnkl}\dt_{mn}\otimes \dt_{kl}=0.
\end{equation}
In this form the section condition implies existence of trivial generalized transformation given by 
$\L_0^{mn}=\e^{mnklp}\dt_{kl}\x_{p}$, i.e. $\d_{\L_0} V^{M}=0$ up to the section condition.

The $E$-bracket is introduced in the usual way via commutation of generalized Lie derivative 
\begin{equation}
[\mc{L}_{\L_1},\mc{L}_{\L_2}]=\mc{L}_{[\L_1,\L_2]_E},
\end{equation}
and reads
\begin{equation}
\label{brackets}
[\L_1,\L_2]_E=[\L_1,\L_2]_D-\fr12Y^{MN}_{KL}\dt_N(\L_1^K\L_2^L).
\end{equation}
Hence, the E-bracket is antisymmetric while the Dorfman bracket is not. Finally, we mention 
following Jacobi identity for the $E$-bracket 
\begin{equation}
\label{Jac_E}
[[\L_{[1},\L_2]_E,\L_{3]}]_E^M=\fr16Y^{MN}_{KL}\dt_N([\L_{[1},\L_2]_E^K\L_{3]}^L).
\end{equation}
This failure of the Jacobi identity and lack of antisymmetric property of the $D$-bracket naturally 
leads to tensor hierarchy in EFT. In other words, tensor fields of higher ranks naturally appear to 
preserve covariance of expressions.

The long space-time derivative, covariant with respect to the D-bracket is defined in the usual way
\begin{equation}
\mc{D}_\m=\dt_\m-\mc{L}_{A_\m}=\dt_\m - \left[A_\m,\bullet\ \right]_D,
\end{equation}
with the generalized vector field $A_\m^M$ transforming as
\begin{equation}
\label{connection}
\d_\L A_\m^M=\dt_\m \L^M-[A_\m,\L]_D^M=\mc{D}_\m\L^M.
\end{equation}
Note, that since D- and E-brackets differ by a trivial transformation (see \eqref{brackets})  the 
above choice is matter of convention. The transformation in this form is taken to keep 
analogy with the conventional Yang-Mills construction.

As usual, the commutator of covariant derivatives defines the field strength of the gauge field 
that fails to be covariant, so one introduces a 2-form gauge field $B_{\m\n}{}^{KL}$ whose degrees 
of freedom are identified with those of the 2-form field $B_{\m\n m}$ via
\begin{equation}
 B_{\m\n m}= 8\e_{mMN}B_{\m\n}{}^{KL}=2\e_{mpqrs}B_{\m\n}{}^{pq\,rs}.
\end{equation}
As a result one has the following tensor hierarchy \cite{Abzalov:2015ega}
\begin{equation}
\label{hierarchy}
\begin{aligned}
[\mc{D}_\m,\mc{D}_\n] =&-\mc{L}_{\F_{\m\n}},\\
\F_{\m\n}^M =&\ 2\dt_{[\m}A_{\n]}^M-[A_{\m},A_{\n}]_E^M+Y^{MN}_{KL}\dt_N B_{\m\n}^{KL},\\
\F_{\m\n\r}{}^{KL}  = &\ 3\,\mc{D}_{[\m}B_{\n\r]}{}^{KL}+
\fr{3}{D(1-2\b_d)}\,Y{}^{KL}_{PQ}\Big(A_{[\m}^{(P}\dt_\n
A{}^{Q)}_{\r]}-\fr13[A_{[\m},A_{\n}]_E{}^{(P}A_{\r]}{}^{Q)}\Big)\\
&-3\big(\dt_N C_{\m\n\r}{}^{N,KL} - Y{}^{KL}_{PQ}\,\dt_N C_{\m\n\r}{}^{Q,PN}\big),\\
\F_{\m\n\r\s}{}^{M,KL}=&\ 
4\,\mc{D}_{[\m}C_{\n\r\s]}{}^{M,KL}+\left(2B_{[\m\n}{}^{KL}\F_{\r\s]}{}^{M}- 
B_{[\m\n}{}^{KL}Y{}^{MN}_{PQ}\dt_NB_{\r\s]}{}^{PQ}\right)\\
&+\fr{4}{3D(1-2\b_d)} 
Y{}^{KL}_{PQ}\left(A_{[\m}^M A_\n^P \dt_\r A_{\s]}^Q - \fr14 A_{[\m}^M [A_\n,A_\r]_E{}^P 
A_{\s]}^Q\right).
\end{aligned}
\end{equation}
The higher rank field strengths are related to the lower rank field strengths via the following 
Bianchi identities
\begin{equation}
\label{bianchi}
\begin{aligned}
3\,\mc{D}_{[\m}\mc{F}_{\n\r]}{}^M= &-Y{}^{MN}_{KL}\dt_N \F_{\m\n\r}{}^{KL},\\
4\,\mc{D}_{[\m}\F_{\n\r\s]}{}^{KL}=&\fr{1}{2}Y{}^{KL}_{PQ}\,\F_{[\m\n}{}^P\F_{\r\s]}{}
{}^Q-3\left(\dt_N
\F_{\m\n\r\s}{}^{N,KL}-Y{}^{KL}_{PQ}\,\dt_N \F_{\m\n\r\s}{}^{Q,PN}\right),\\
 5\mc{D}_\m \mc{F}_{\n\r\s\k}^{N,KL}=&\fr{10}{3} \mc{F}_{[\m\n}^{N}\mc{F}_{\r\s\k]}{}^{KL}+\ldots,
\end{aligned}
\end{equation}
where the dots in the last line denote terms that always drop from covariant expressions below 
because of the section condition.

Under arbitrary variations of the $p$-form potentials the covariant field strengths transform as 
follows:
\begin{equation}
\label{varF}
\begin{aligned}
\d \F_{\m\n}{}^M=&\ 2\,\mc{D}_{[\m}\D A_{\n]}^M-Y{}^{MN}_{KL}\dt_N \D B_{\m\n}{}^{KL},\\
\d \F_{\m\n\r}{}^{KL}=&\ 3\,\mc{D}_{[\m}\D 
B_{\n\r]}{}^{KL}+\fr{1}{2}Y{}^{KL}_{PQ}\,\F_{[\m\n}{}^P\D 
A_{\r]}^Q \\
&-3\big( \dt_N \D C_{\m\n\r}{}^{N,KL} - Y{}^{KL}_{PQ}\, \dt_N \D C_{\m\n\r}{}^{Q,PN}\big),\\
\d \F_{\m\n\r\s}{}^{M,KL}=&\ 4\, D_{[\m}\D C_{\n\r\s]}{}^{M,KL}+\fr{1}{18} 
\Big(\fr{3}{8}\F_{[\m\n}{}^{M}\D B_{\r\s]}{}^{KL}-\fr{1}{4}\F_{[\m\n\r}{}^{KL}\d A_{\s ]}^M\Big)
\end{aligned}
\end{equation}
where it proves useful to define ``covariant'' transformations 
\begin{equation}
\begin{aligned}
\D A_\m^M=&\ \d A_\m^M,\\
\D B_{\m\n}{}^{KL}=&\ \d B_{\m\n}{}^{KL}-\fr{1}{6}Y{}^{KL}_{MN}A_{[\m}^{M}\d A_{\n]}^{N},\\
\D C_{\m\n\r}{}^{N,KL} =&\  \d C_{\m\n\r}{}^{N,KL} - \d A_{[\m}^N B_{\n\r]}{}^{KL} - 
\fr{1}{18}
Y{}^{KL}_{RS} A_{[\m}^N A_\n^R \d A_{\r]}^S.
\end{aligned}
\end{equation}
For the gauge transformations this gives
\begin{equation}
\label{trans_AB}
\begin{aligned}
\D A_\m^M =&\ \mc{D}_\m\L^M+Y{}^{MN}_{KL}\dt_N\X_\m{}^{KL},\\
\D B_{\m\n}{}^{KL} =& \
2\mc{D}_{[\m}\X_{\n]}{}^{KL}-\fr{1}{6}Y{}^{KL}_{MN}\L^M\mc{F}_{\m\n}{}^N+3\left(\dt_N\Y_{\m\n}{}^{N
,KL}-Y{}^{KL}_{PQ}\dt_N \Y_{\m\n}{}^{P,NQ}\right),\\
\D C_{\m\n\r}{}^{M,KL}=&\ 3\mc{D}_{[\m} 
\Y_{\n\r]}{}^{M,KL}-\mc{F}_{[\m\n}{}^M\X_{\r]}{}^{KL}+\fr{1}{9}Y{}^{KL}_{PQ}\L^P\mc{F}_{
\m\n\r}{}^
{
QM}.
\end{aligned}
\end{equation}
The above transformations are constructed such that the covariant 2-, 3- and 4-form field 
strengths are indeed covariant with respect to $\L^M$, $\X_\m{}^{KL}$ and $\Y_{\m\n}{}^{N,KL}$ 
transformations.

Since the 2- and 3-forms above are related to the ones parameterizing the supergravity degrees 
of freedom and duals used in \cite{Samtleben:2005bp} as
\begin{equation}
\label{proper_fields}
 \begin{aligned}
   B_{\m\n i}&=2\e_{iklmn}B_{\m\n}{}^{klmn},\\
   C_{\m\n\r}{}^m&=-6\e_{nklrs}C_{\m\n\r}{}^{mn,klrs},
 \end{aligned}
\end{equation}
it is convenient to  rewrite the covariant transformation as
\begin{equation}
\begin{aligned}
 \D A_\m^{mn}=&\ \d A_\m^{mn},\\
\D B_{\m\n i}=&\ \d B_{\m\n i}-2\e_{imnkl}A_{[\m}^{mn}\d A_{\n]}^{kl},\\
\D C_{\m\n\r}{}^{m} =&\  \d C_{\m\n\r}{}^{m} + 3\d A_{[\m}^{mn} B_{\n\r] n} -2 \e_{nklrs} 
A_{[\m}^{mn} A_\n^{kl} \d A_{\r]}^{rs}.
\end{aligned}
\end{equation}
Here one should take into account the factor 1/2, that is necessary to prevent double 
counting when going from capital Latin indices  to double small indices in a contraction. 
With the fields defined in \eqref{proper_fields} and the corresponding relation for the gauge 
parameters the gauge 
transformations read
\begin{equation}
 \begin{aligned}
  \D A_\m^{mn} =&\ \mc{D}_\m\L^{mn}+\fr{1}{16}\e^{imnkl}\dt_{kl}\X_{\m i},\\
\D B_{\m\n i} =& \
2\mc{D}_{[\m}\X_{\n] 
i}-2\e_{imnpq}\L^{mn}\mc{F}_{\m\n}{}^{pq}-\dt_{mi}\Y_{\m\n}{}^{m},\\
\D C_{\m\n\r}{}^m=&\ 3\mc{D}_{[\m} 
\Y_{\n\r]}{}^{m}+3\mc{F}_{[\m\n}{}^{mn}\X_{\r]n}+\L^{mn}\mc{F}_{
\m\n\r n},
 \end{aligned}
\end{equation}
where the identity \eqref{use_id1} has been used. 
Note that these have precisely the same form is in the $D=7$ maximal gauged supergravity up to the 
following mnemonic replacements of derivatives along extended coordinates by components $Y_{mn}$ and $Z^{mn,k}$ of 
embedding tensor.
\begin{equation}
\begin{aligned}
  \e^{imnkl}\dt_{kl}V_{i}&=-16gZ^{mn,i}V_i \\
  \dt_{mn}V^{m}&=\fr{g}{24}Y_{mn}V^m.
\end{aligned}
\end{equation}
Certainly, the correct way to check that the transformations indeed match is to perform Scherk-Schwarz reduction explicitly, possibly, dropping the trombone gauging. Although being an interesting project by itself, this is beyond the scope of the present work.

The same is true for the Bianchi identities that for the fields \eqref{proper_fields} take the 
following nice form
\begin{equation}
 \begin{aligned}
   3\DD_{[\m} \F_{\n\r]}{}^{mn}&=-\fr{1}{16}\e^{imnkl}\dt_{kl}\F_{\m\n\r i},\\
   4\DD_{[\m}\F_{\n\r\s m} 
    &=6\e_{mpqrs}\F_{[\m\n}{}^{pq}\F_{\r\s]}{}^{rs}+\dt_{nm}\F_{\m\n\r\s}{}^{n},\\
   5\DD_{[\m}\F_{\n\r\s\k]}{}^m&=-10\F_{[\m\n}{}^{mn}\F_{\r\s\k n}+\ldots 
 \end{aligned}
\end{equation}
In the non-coordinate notation the above equations read
\begin{equation}
 \begin{aligned}
   \DD\F^{mn}&=\fr{1}{16}\e^{imnkl}\dt_{kl}\F_i,\\
   \DD\F_m&=\e_{mpqrs}\F^{pq}\wedge \F^{rs}+\dt_{nm}\F^n,\\
   \DD\F^m&=-\fr15 \F^{mn}\wedge \F_n+\ldots,
 \end{aligned}
\end{equation}
where we define a $p$-form $\w_p$ in terms of its components in the usual way
\begin{equation}
 \w_p=\fr{1}{p!}\w_{i_1\ldots i_p}dx^{i_1}\wedge \cdots \wedge dx^{i_p}.
\end{equation}

\section{Covariant exceptional field theory}
\label{EFT}

The full SL(5)-covariant Exceptional Field Theory Lagrangian has the following structure 
\begin{equation}
\begin{aligned}
\LL_{EFT}=&\ \LL_{EH}(\hat R)+\LL_{sc}(\mc{D}_\m 
m_{kl})+\LL_{V}(\F_{\m\n}{}^{mn})+\LL_{T}(\F_{\m\n\r\, m})\\
&+\LL_{top}-V(m_{kl},g_{\m\n}).
\end{aligned}
\end{equation}
Here the modified Einstein-Hilbert term $\LL_{EH}$  written in terms of the modified curvature, the 
kinetic terms for the scalar fields $\LL_{sc}$ and for the vectors $\LL_{V}$ and the invariant 
potential for the scalar fields $V$ have the same structural form as in other EFT's, see for example  \cite{Hohm:2013jma,Hohm:2013pua,Hohm:2013vpa} for the modified EH term and \cite{Abzalov:2015ega} for the general form of the potential.

The kinetic term for the 2-form gauge potentials $B_{\m\n n}$ appears here as a proper Lagrangian 
governing dynamics of the corresponding degrees of freedom. Whereas, in the SO(5,5) theory such a 
term was subject to a self-duality condition and for higher rank U-duality groups was not there at 
all.

Finally, the topological term has always different structure depending on the dimension and the duality group 
and hence has to be processed separately. While U-duality covariance of the other terms is 
explicit, the topological term does not have a covariant form. Instead, one may write its variation in a covariant form, that is the only relevant expression to recover EOM's.

We should mention here, that 6+1-dimensional diffeomorphisms for the scalar and vector kinetic 
terms, the modified Einstein-Hilbert term and the scalar potential work precisely as in 
\cite{Hohm:2013vpa,Abzalov:2015ega} 
and hence invariance is  not explicitly checked here. However, we perform explicit check of invariance of of the 2-form kinetic term and the topological term with respect to external 
diffeomorphisms, that successfully 
fixes \emph{all} the coefficients in the Lagrangian. This is a known feature of EFT in contrast to the 
maximal gauged supergravity, where all the coefficients become fixed only after imposing 
supersymmetry condition. One may speculate that already the bosonic EFT contains some information about the full supersymmetric theory.

\subsection{Kinetic Lagrangian and invariant potential}

The fully covariant Einstein-Hilbert term takes the following usual form
\begin{equation}
S_{EH}=-\fr12\int d^n x d^D\XX e \hat{R}= -\fr12\int d^n x d^D\XX e e^\m_a e^\n_b 
\hat{R}_{\m\n}{}^{ab},
\end{equation}
where the modified curvature reads
\begin{equation}
\hat{R}_{\m\n ab}=R_{\m\n ab}+\F_{\m\n}{}^Me^\r_a\dt_M e_{\r b}.
\end{equation}
To ensure invariance of the Einstein-Hilbert term with respect to local Lorentz transformations 
depending on extended coordinates, the corresponding spin-connection 
$\w_\m{}^{ab}$ is set to have weight zero. One should consider this general dependence of local 
transformations since all fields in the theory depend on the extended coordinates.

The corresponding Lorentz-invariant Riemann scalar then differs 
from the usual expression and has the same form as in \cite{Hohm:2013vpa}.
The usual equation that determines the spin-connection 
can be written in the following covariant form
\begin{equation}
\mc{D}_{[\m} e_{\n]}^a-\fr14 \w_{[\m}{}^{ab}e_{\n]b}=0.
\end{equation}
As was checked in \cite{Abzalov:2015ega} in general form invariance of the scalar potential implies, that the external vielbein is a generalized scalar of weight $\l(e^a_\m)=1/5$.

For the scalar degrees of freedom parametrized by the matrix $\mc{M}_{MN}$ we just use the general 
result  and set $\a_d=3$ \cite{Abzalov:2015ega}
\begin{equation}
\begin{aligned}
\LL_{sc}&=\fr{1}{12} eg^{\m\n}\mc{D}_\m \M_{MN}\mc{D}_\n \M^{MN},\\
\M_{mn\,kl}&=m_{mk}m_{nl}-m_{ml}m_{nk}.
\end{aligned}
\end{equation}
As expected this  is explicitly covariant under the local gauge transformation 
generated by the 
generalized Lie derivative. 

Finally, the kinetic terms for the 1-form potential $A_\m^M$ and the 2-form potential $B_{\m\n m}$ 
take the following form
\begin{equation}
\begin{aligned}
\LL_{V}&=-\fr14 e\M_{MN}\F_{\m\n}{}^M\F_{\m\n}{}^N=-\fr{1}{8}em_{mk}m_{nl}\F_{\m\n}{}^{mn}\F^{\m\n 
kl},\\
\LL_{T}&=-\fr{1}{3\cdot(16)^2}em^{mn}\F_{\m\n\r\, m}\F^{\m\n\r}{}_{n}
\end{aligned}
\end{equation}
Each term here is separately covariant with respect to generalized diffeomorphisms and all gauge 
transformations. It is explicitly shown further, that the numerical coefficient in $\LL_{T}$ is defined by invariance under 6+1-dimensional external diffeomorphisms.

The non-topological part of the Lagrangian is concluded by the so-called scalar potential. This 
depends only on the scalar and metric degrees of freedom and their derivatives with respect to 
extended coordinates $\XX^{mn}$. This has a form universal for the duality groups $E_{6,5}$, 
SO(5,5), SL(5) and SL(2)$\times$SL(3):
\begin{equation}
\label{V}
\begin{aligned}
V=&\ -\fr{1}{4\a_d}\M^{MN}\dt_M \M^{KL}\dt_N \M_{KL}+\fr12 \M^{MN}\dt_M \M^{KL}\dt_L \M_{NK}\\
&-\fr12 (g^{-1}\dt_M g)\dt_N \M^{MN}-\fr14 \M^{MN}(g^{-1}\dt_M g)(g^{-1}\dt_N g)-\fr14 \M^{MN}\dt_M 
g^{\m\n}\dt_{N}g_{\m\n},
\end{aligned}
\end{equation}
where the terms in the first line are precisely those of \cite{Berman:2011jh} and the rest terms 
are 
needed to ensure gauge invariance, and one should note the determinant $\sqrt{-g}$ in the action. 
For the case in question one sets $\a_d=3$. Covariance of the above expression has been explicitly 
checked in \cite{Abzalov:2015ega}.

\subsection{Topological Lagrangian}

To construct the topological term one notes that in the embedding tensor formulation of the 
maximal $D=7$ supergravity the field $C_{\m\n\r}{}^m$ appears only under projection with the tensor 
\begin{equation}
Y_{mn}=\fr12\Q_{p(m,n)}{}^p,
\end{equation} 
that parametrizes gaugings in the $\bf 15$ of SL(5). Analyzing the expressions for  the 3-form 
field strength $\F_{\m\n\r}{}^{KL}$ and for the covariant transformation $\D B_{\m\n}{}^{KL}$ one 
arrives to the following rule
\begin{equation}
\begin{aligned}
Y_{mn}C_{\m\n\r}{}^n && \to && \e_{mKL}\Big(\dt_N
C_{\m\n\r}^{N,KL}-Y^{KL}_{PQ}\dt_N C_{\m\n\r}^{Q,PN}\Big).
\end{aligned}
\end{equation}
Indeed, since the field $C_{\m\n\r}{}^{mn,klrs}$ contains only the representation $\bf 5$ of SL(5) one has the 
identity $12\e_{pklrs}C_{\m\n\r}{}^{mn,klrs}=C_{\m\n\r}{}^{[m}\d^{n]}_p$.
Now performing Scherk-Schwarz reduction 
$C_{\m\n\r}{}^r(x^\m,\XX^M)=V^r_{\bar{r}}(\XX^M)C^{\bar{r}}(x^\m)$ one obtains
\begin{equation}
\begin{aligned}
\e_{iKL}\Big(\dt_N
C_{\m\n\r}^{N,KL}-Y^{KL}_{PQ}\dt_N C_{\m\n\r}^{Q,PN}\Big) && \to && 
(V_{\bp}^p\dt_{pq}V^q_{\bq})C_{\m\n\r}{}^{\bq}.
\end{aligned}
\end{equation}
Using the SL(5) gaugings written in terms of twist matrices obtained in \cite{Berman:2012uy} the 
expression in brackets on the RHS becomes precisely $Y_{\bp \bq}$ (for the vanishing trombone 
gauging and $\det V=1$). Barred indices are used only in the paragraph above and denote the flat directions of the Scherk-Schwarz twist matrix, see \cite{Berman:2012uy,Musaev:2013rq}. 

Hence, inspired by these observations, we write variation of the topological Lagrangian in the 
following simple form (cf. 
\cite{Samtleben:2005bp})
\begin{multline}
\d \mc{L}_{top}= A\e^{\m\n\r\l\s\t\k}\bigg[\F_{\m\n\r\l}{}^{i}\e_{iKL}\Big(\dt_N
\D C_{\s\t\k}^{N,KL}-Y^{KL}_{PQ}\dt_N \D C_{\s\t\k}^{Q,PN}\Big)\\
+\fr14\F_{\m\n}{}^{ij}\F_{\r\l\s i}\D B_{\t\k j}-\fr{1}{12}\F_{\m\n\r i}\F_{\l\s\t j}\d 
A_\k^{ij}\bigg] 
+\mbox{total derivatives}.
\end{multline}
Here the coefficients are chosen for the variation to vanish on all gauge and U-duality 
transformation. Since it is easier to work with the fields \eqref{proper_fields} we rewrite  the above 
expression as
\begin{equation}
\label{var_top}
 \begin{aligned}
  \d \mc{L}_{top}= A\e^{\m\n\r\l\s\t\k}\bigg[\F_{\m\n\r\l}{}^{i}\dt_{ij} \D C_{\s\t\k}^{j}
+6\F_{\m\n}{}^{ij}\F_{\r\l\s i}\D B_{\t\k j}-2\F_{\m\n\r i}\F_{\l\s\t j}\d 
A_\k^{ij}\bigg], 
 \end{aligned}
\end{equation}
where the overall prefactor will be fixed to $A^{-1}=16\cdot 4! $ by invariance with respect to 
6+1-dimensional external diffeomorphisms. Note, that $\e^{\m\n\r\s\l\t\k}$ here and always in the 
paper is the alternating symbol, rather than the Levi-Civita tensor, and hence does not contain the 
determinant $e$.

As for the topological terms of EFT's in other dimensions (as well as of gauged supergravities) 
this expression cannot be written as variation of a covariant expression. However, the above is 
enough to write equations of motion and to check invariance of the Lagrangian with respect to 
6+1-dimensional diffeomorphisms.

\subsection{$D=6+1$ diffeomorphisms}

In the previous section we have established the explicit form of the kinetic Lagrangian for the 
fields $A_\m{}^{mn}$, $B_{\m\n m}$ and $\M^{MN}$, the modified Einstein-Hilbert term, the 
scalar potential $V(\dt\M,\M)$ and the topological term. These are invariant under duality transformations
as well as under all the gauge transformations resulting from the tensor hierarchy. This 
invariance fixed for us all the mutual prefactors in the Lagrangian except the prefactor of the 
topological term and the kinetic term of the 2-form gauge potential. It is known, that the same situation appears in the maximal gauged supergravity 
models, where to fix the remaining prefactor on needs to consider supersymmetry. 

The case of Exceptional Field Theory is different due to dependence of all the fields on the extra coordinates 
$\XX^M$. This results in the fact, that external 6+1-dimensional diffeomorphisms do not 
work automatically and one has to perform a certain check of that. Remarkably, it is enough just to 
fix these two prefactors to satisfy the invariance condition. The result is a completely fixed duality, 
gauge and external diffeomorphism invariant Lagrangian. Another miracle appears, when one checks 
supersymmetry of the (SUSY extended) Lagrangian and gets that for free. This has been checked 
explicitly for the duality groups $E_{7,6}$ in \cite{Godazgar:2014nqa} and \cite{Musaev:2014lna}, 
however there is no reason to expect, that other U-duality groups fail to follow this scheme.

Hence, let us start with the following external diffeomorphism transformations
\begin{equation}
 \begin{aligned}
  \d e^a_\m &=\x^\m \mc{D}_\n e^a_\m+\mc{D}_\m \x^\n e^a_\n,\\
  \d \M_{MN} &= \x^\m \mc{D}_\m \M_{MN},\\
   \d A_\m{}^{M} &= \x^\n \F_{\n\m}{}^{M}+ \M^{MN}g_{\m\n}\dt_N \x^\n,\\
     \D B_{\m\n i}&=\x^\r \mc{F}_{\r\m\n i},\\
   \D C_{\m\n\r}{}^{m}&= -\fr{1}{3!}e\e_{\m\n\r\s\k\l\t}\x^\s m^{mn}\F^{\k\l\t}{}_{n},
 \end{aligned}
\end{equation}
where $\ve_{\m\n\r\s\k\l\t}=e\e_{\m\n\r\s\k\l\t}$ is the Levi-Civita tensor in 7 dimensions.
Transformation of the 3-form potential is required to be of this particular form by 
\emph{off-shell} invariance of the Lagrangian (see remark at the end of this section). 

One should note, that huge part of cancellations here works precisely as in the maximal gauged 
supergravity and hence, does not need to be double checked. In contrast, the terms that contain 
the derivative $\dt_{mn} \x^\m$ do not exist in the gauged models and hence need to be processed 
explicitly. We will refer to them as \emph{new terms} and work in the close analogy to 
\cite{Hohm:2013vpa}. Next we note that diffeomorphism invariance of the universal 
scalar potential 
$V$ has been checked in general form in \cite{Abzalov:2015ega}, hence we just use the result here.

Let us start with variations of the 2- and 3-form field strengths and write for the former
\begin{equation}
\label{dF2}
 \begin{aligned}
  \d \F_{\m\n}{}^{mn}=&\ 2\DD_{[\m}\D A_{\n]}{}^{mn}-\fr{1}{16}\e^{mnpqr}\dt_{pq} B_{\m\n r}\\
  =&\ 2 \DD_{[\m} (\x^\r \F_{\r\m}{}^{mn})+\DD_{[\m}(\M^{mn,pq}g_{\n]\r}\dt_{pq} 
\x^\r) -\fr{1}{16}\e^{mnpqr}\dt_{pq}\dt_{pq}(\x^\r \F_{\m\n\r r})\\
=& \ (L_\x^\DD \F_{\m\n}{}^{mn})-\fr{1}{16}\e^{mnpqr}\F_{\m\n\r r}\dt_{pq} 
\x^\r +2\DD_{[\m}(m^{mp}m^{nq}g_{\n]\r}\dt_{pq} \x^\r),
 \end{aligned}
\end{equation}
where we used the Bianchi identity for the field $\F_{\m\n}{}^{mn}$ to organize the (conventional) 
Lie derivative with respect to $\DD_\m$ that is denoted by $L_\x^\DD$.  Using the same arguments 
and the Bianchi identity for the 3-form field strength we write for its variation
\begin{equation}
\label{dF3}
 \begin{aligned}
  \d \F_{\m\n\r m}=&\  3\DD_{[\m}\D B_{\n\r]m}+6\e_{mpqrs}\F_{[\m\n}{}^{pq}\d 
A_{\r]}{}^{rs}-\dt_{mn}\D C_{\m\n\r}{}^{n}\\
   =&\ 3\DD_{[\m}(\x^\s\F_{\n\r\s] m})+6\e_{mpqrs}\F_{[\m\n}{}^{pq}\F_{\s\r]}{}^{rs} \x^\s 
+3\e_{mpqrs}\F_{[\m\n}{}^{pq}\M^{rs,kl}g_{\r\s}\dt_{kl} \x^\s\\
&-\dt_{mn}\D C_{\m\n\r}{}^{n}\\
=&\ (L^\DD_\x \F_{\m\n\r 
m})+6\e_{mpqrs}m^{rk}m^{sl}\F_{\m\n}{}^{pq}g_{\r\s}\dt_{kl}\x^\s+\x^\s\dt_{mn}\F_{\m\n\r\s}{}^{n}
-\dt_{mn}\D C_{\m\n\r}{}^{n}.
 \end{aligned}
\end{equation}
Here the transformation of the 3-form gauge potential $\D C_{\m\n\r}{}^m$ was left inexplicit for 
further convenience.

Now, one notices that terms in the variation of the Lagrangian containing the (conventional) Lie 
derivative of the 2- and 3-form potentials together with variations of the determinant of the 
vielbein $\det e$ and the scalar matrix $m^{mn}$ in the corresponding kinetic terms give just full 
derivative. This is exactly the same as in the gauged theories and in other EFT's. Next, from the 
analysis of EFT's for the other U-duality groups one concludes 
that the last term in 
the last line of \eqref{dF2} will cancel against the corresponding contribution from the variation 
of the modified Einstein-Hilbert term.

Hence, what is left are the following six terms
\begin{equation}
 \begin{aligned}
   (1)&= \fr{1}{4\cdot 16}e m_{mk}m_{nl}\e^{mnpqr}\F^{\m\n kl}\F_{\r\m\n r}\dt_{pq}\x^\r,\\
   (2)&=-\fr{1}{4\cdot 16}e 
m^{mn}\e_{mpqrs}m^{rk}m^{sl}\F_{\m\n}{}^{pq}\F^{\m\n\r}{}_{n}g_{\r\s}\dt_{kl}\x^\s,\\
(3)&=\fr{1}{16}em^{mn}\F^{\m\n\r}{}_n\Big(\x^\s\dt_{mk}\F_{\m\n\r\s}{}^k-\dt_{mk}\D 
C_{\m\n\r}{}^k\Big),\\
(4)&=A\e^{\m\n\r\l\s\t\k}\F_{\m\n\r\l}{}^m\dt_{mn}\D C_{\s\t\k}^{n},\\
(5)&=-2A\e^{\m\n\r\l\s\t\k}\F_{\m\n\r m}\F_{\l\s\t 
n}\big(\x^\y\F_{\y\k}{}^{mn}+g_{\k\y}m^{mk}m^{nl}\dt_{kl}\x^{\y}\big),\\
(6)&=6A\e^{\m\n\r\l\s\t\k}\F_{\m\n}{}^{mn}\F_{\r\l\s m}\F_{\y\t\k 
n}\x^\y
 \end{aligned}
\end{equation}
where the first line comes from the variation of the kinetic term for the 1-form 
gauge potential, the second and third lines comes from the kinetic term for the 2-form gauge 
potential. The last three lines result from the expression  \eqref{var_top}, with the line (5) 
resulting from the term with $\d A_\m{}^{mn}$ and the line (6) from the term with $\D B_{\m\n m}$. 
One immediately notes here, that the first term in brackets in the line (5) together with the line 
(6) forms an expression with the 8 indices $\{\m\n\r\l\s\t\k\y\}$ fully antisymmetrized, and hence 
cancels.

Next, the lines (1) and (2) cancel against each other since the scalar matrix $m\in$ SL(5). Indeed, 
as a consequence of $\det m=1$ we may write
\begin{equation}
 m^{mn}m^{rk}m^{sl}\e_{mpqrs}=\e^{nklij}m_{ip}m_{jq},
\end{equation}
that after substituting into (2) gives the desired cancellation. Note, that the identity above can be 
understood as a rule for raising and lowering the indices of the alternating symbol, however we 
will not need this.

Now, we note that the first term [3.1] in the line (3) and the term in the line (4) can be combined 
into a full derivative. Indeed, using the transformation of the 3-form gauge potential we write
\begin{equation}
\begin{aligned}
 &[3.1]+(4)=\\
 &=-\fr{1}{16}e 
m^{nk}\F^{\m\n\r}{}_{k}\x^\s\dt_{nm}\F_{\m\n\r\s}{}^m-\fr{A}{3!}\e^{\m\n\r\l\s\t\k}\F_{\m\n\r\l}{}
^m\dt_{mn}\big(e\e_{\s\t\k\y\c\t\w}\x^\y\F^{\c\t\w}{}_{k}m^{nk}\big)\\
 &=-\fr{1}{16}e m^{nk}\F^{\m\n\r}{}_{k}\x^\s\dt_{nm}\F_{\m\n\r\s}{}^m -
 A\cdot 4!\F_{\m\n\r\s}{}^m\dt_{mn}\big(e \x^\m \F^{\n\r\s}{}_{k}m^{nk}\big).
 \end{aligned}
\end{equation}
Setting the prefactor $A^{-1}=16\cdot4!$ and taking into account two minus signs resulting from 
necessary permutation, one arrives to a full derivative, that drops from the variation of the 
action. 

With the prefactor $A$ being fixed we are left with check of the cancellation between the 
second terms in the lines (3) and (5). This is straightforward, however one should take care of the 
$\det e$ prefactors. Hence, we write (multiplied by $(32 \cdot 3!)$ for convenience)
\begin{equation}
 \begin{aligned}
   &(32 \cdot 3!)\Big([3.2]+[5.2]\Big)=\\
   =&-2em^{mn}\F^{\m\n\r}{}_{n}\dt_{mk}\big(e \e_{\m\n\r\s\k\l\t}\x^\s\F^{\k\l\t}{}_{l}m^{kl}\big)  
-\e^{\m\n\r\l\s\t\k}\F_{\m\n\r m}\F_{\l\s\t n}g_{\k\y}m^{mk}m^{nl}\dt_{kl}\x^{\y}\\
=&-2 \e_{\m\n\r\s\k\l\t}\dt_{mk}\big(em^{mn}\F^{\m\n\r}{}_{n}\big) e\F^{\k\l\t}{}_{l}m^{kl} \x^\s 
-e \cdot \ve_{\m\n\r\l\s\t\k}\F^{\m\n\r}{}_{m}\F^{\l\s\t}{}_{n}m^{mk}m^{nl}\dt_{kl}\x^{\k}\\
=&-\e_{\m\n\r\s\k\l\t}\dt_{mk}\big(e^2m^{mn}\F^{\m\n\r}{}_{n}\F^{\k\l\t}{}_{l}m^{kl} \x^\s\big) 
\Rightarrow0.
 \end{aligned}
\end{equation}
Note the use of the Levi-Civita tensor $\ve_{\m\n\r\l\s\t\k}$ in the second expression, that 
produces an extra factor of $e$ times the (constant) alternating symbol.

As the final remark in this section let us look at the equations of motion for the (non-dynamical) 
field $C_{\m\n\r}{}^m$, that read
\begin{equation}
 \dt_{mk}\big(e m^{mn}\F^{\m\n\r}{}_{n}-\fr{1}{4!}\e^{\m\n\r\l\s\t\k}\F_{\l\s\t\k}{}^m)=0.
\end{equation}
The result is that the 3-form gauge potential does not give dynamical field equations in the 
external 6+1-dimensional space-time. Rather, it results in restricting of the 3- and 4-form field 
strength behavior in the internal extended space. After Scherk-Schwarz reduction the above 
equation results in the known duality relation between the 3- and 4-form field strengths. This is 
an expected result, as the fifth component of the 2-form gauge potential was introduced as a 
dualization of the 11-dimensional 3-form gauge field with all 
indices external. Since one was always allowed to dualize the 2-form gauge degrees of freedom to 
get a 3-form gauge potential, to keep the story duality covariant one should introduce both the 2- 
and the 3-form gauge potentials. This doubling of fields is the price for having the theory duality 
covariant. The final field content of the model depends on the  gauging chosen.

Finally, let us note, that upon imposing the following dualization constraint
\begin{equation}
 e m^{mn}\F^{\m\n\r}{}_{n}-\fr{1}{4!}\e^{\m\n\r\l\s\t\k}\F_{\l\s\t\k}{}^m,
\end{equation}
the diffeomorphism transformation rule for the 3-form field strength takes its conventional form
\begin{equation}
 \D_\x C_{\m\n\r}{}^m=\x^\s\F_{\s\m\n\r}{}^m.
\end{equation}

\section{Embeddings of D=11 and Type IIB supergravity}
\label{Embed}

The field content of the 11-dimensional and Type IIB supergravity can be naturally embedded into the 
field content of the exceptional field theory upon a correct choice of the solution of the section 
condition. Depending on the duality group one gets a different splitting of the 
coordinates of the resulting theory. As was shown in \cite{Hohm:2013vpa} for the E${}_6$ exceptional 
field theory the resulting Lagrangian  does not preserve the full D=10 Lorenz invariance due to this	 
coordinate split. Since there is no reason to expect that on the level of the Lagrangian the construction works only for  the 
E${}_6$ duality group, where it has been checked explicitly, we perform here only the  check of the field 
content. However, in principle one would be interested in having an explicit picture of how the 
Lagrangian of all the EFT's reduces to the known supergravities.

Let us start with embedding of the 11-dimensional supergravity field content. The corresponding 
solution of the section condition breaks the U-duality group SL(5) to GL(4) and provides the 
following decomposition of the relevant representations
\begin{equation}
\begin{aligned}
&& SL(5) & \to SL(4)\times GL(1) \simeq GL(4); \\
&& \bf 10 & \to {\bf 4}_{-3}+{\bf 6}_{2};\\
&& \bf 5 & \to {\bf 4}_{1}+{\bf 1}_{-4};\\
&& \bf 24 & \to {\bf 1}_{0}+{\bf 4}_{5}+{\bf \bar{4}}_{-5}+{\bf 15}_0,
\end{aligned}
\end{equation}
where the subscripts denote weights with respect to the $GL(1)$ subgroup. Since the extended 
coordinates $\XX^{mn}$ transform under the representation $\bf 10$ they decompose according to the 
second line above, that gives
\begin{equation}
\XX^{mn} \to \{\XX^{5 a}, \XX^{ab}\} \to \{x^a, \e^{abcd}y_{cd}\},
\end{equation}
where $\e^{abcd}$ is the alternating symbol in 4 dimensions. The coordinates $x^a$ have the 
interpretation of the usual geometric coordinates, while $y_{ab}$ correspond to winding modes of the 
M2-brane. It is straightforward to check that dropping dependence on the winding coordinates solves 
the section condition
\begin{equation}
\e^{imnkl}\dt_{mn}\otimes \dt_{kl}=0.
\end{equation}
Hence, all the fields of the theory depend only on eleven coordinates: the space-time external 
coordinates $x^\m$ and the internal ones $x^a$.

The corresponding decomposition of the gauge fields works as follows
\begin{equation}
\begin{aligned}
&& A_\m{}^{mn} & \to A_\m{}^{a},\, A_{\m ab},\\
&& B_{\m\n m} & \to B_{\m\n}, \, B_{\m\n a}.
\end{aligned}
\end{equation}
Here we do not include the field $C_{\m\n\r}{}^m$ as it completely drops from the theory on the 
solution of the section condition. This decomposition nicely fits into the decomposition of the 
metric $G_{\tt\hat{M}\hat{N}}$ and the 3-form field $C_{\tt \hat{M}\hat{N}\hat{K}}$ in 11 dimensions (see 
\eqref{split})
\begin{equation}
\begin{aligned}
&& G_{\tt\hat{M}\hat{N}} & \to \{g_{\m\n},\, A_{\m}{}^{a},\, \f_{ab} \},\\
&& C_{\tt\hat{M}\hat{N}\hat{K}} & \to \{C_{\m\n\r},\,  B_{\m\n a},\,  A_{\m ab},\, \f_{abc} \}
\end{aligned}
\end{equation}
The 3-form field $C_{\m\n\r}$ is obtained by dualization of the 2-form $B_{\m\n}$ in 7 dimensions. 
The 14 scalars $\f_{ab}$ and $\f_{abc}$ is the above decomposition are identified with the 
components of the generalized metric $m_{mn}$ that lives in the $\bf 24$ of SL(5) factorized by the 
$\bf 10$ of SO(5) considered as its subgroup. Hence a combination of $\bf 4_{5}$ and $\bf 
\bar{4}_{-5}$ is factored out as well as the SO(4) part of the $\bf 15_0$. The latter together with 
the singlet $\bf 1$ form the coset space GL(4)/SO(4), while the remained $\bf 4$ gives the fields 
$\f_{abc}$. The easiest way to see this is to look at decomposition in the matrix representation of 
the groups SL(5) and SO(5)
\begin{equation}
\begin{aligned}
&& SL(5): & 
\begin{bmatrix}
SL(4) & \bf 4_5 \\
\bf \bar{4}_{-5} & \bf 1_0
\end{bmatrix},
&& &
SO(5): & 
\begin{bmatrix}
SO(4) & \bf 4 \\
\bf 4 & 1
\end{bmatrix}
\end{aligned}.
\end{equation}

The other possible branching SL(5)$\to$GL(3)$\times$SL(2) gives the field content of 
the ten-dimensional Type IIB supergravity with 7+3 split. As usual for EFT's, the explicit SL(2) 
symmetry is identified with the S-duality symmetry of the theory. Branching rules for the relevant 
representations take the following form
\begin{equation}
 \begin{aligned}
  && \bf 10 & \to  ({\bf 1,1})_{-6} + ({\bf \bar{3},1})_4+({\bf 3,1})_{-1},\\
  && \bf 5 & \to  ({\bf 1,2})_{-3} + ({\bf 3,1})_2,\\
  && \bf 24 & \to  ({\bf 1,1})_0 +({\bf 1,3})_0+({\bf 3,2})_5+ ({\bf \bar{3},2})_{-5}+({\bf 8 
,1})_0,
 \end{aligned}
\end{equation}
where the first irrep corresponds to the SL(3) subgroup of GL(3) and the subscript denotes weight 
with respect to its GL(1) subgroup. The first line above implies the following decomposition of the 
extended coordinates $\XX^{mn}$
\begin{equation}
\label{sl3decomp}
 \XX^{mn}\to \{\XX^{\un{a}\un{b}},\XX^{\un{a}\hat{\a}},\XX^{\hat{a}\hat{\b}}\}\to 
\{\e^{\un{a}\un{b}\un{c}}x_{\un{c}},y^{\un{a},\hat{\a}},\e^{\hat{\a}\hat{\b}}z\},
\end{equation}
where $\e^{\un{a}\un{b}\un{c}}$ and $\e^{\hat{\a}\hat{\b}}$ are the alternating symbols for the 
SL(3) and SL(2) groups respectively Supergravity interpretation of the above decomposition in 
terms of the geometric coordinates and winding modes of various Type II branes needs more careful 
consideration. 

First, one should note that breaking the SL(2) symmetry explicitly and leaving only the 
coordinates $\{x_{\un{a}},y^{un{a},1}\}$ results in the O(d,d) theory, that is the Double Field 
Theory \cite{Thompson:2011uw}. The DFT section condition
\begin{equation}
 \fr{\dt}{\dt x_{\un{a}}}\otimes\fr{\dt}{\dt y^{\un{a}}}=0 
\end{equation}
is a direct consequence of the SL(5) section condition. Hence, as it is known from DFT, to return 
to the Type IIA theory one just drops dependence on the $x_{\un{a}}$ coordinates, that correspond 
to winding of the fundamental string of Type IIA. Alternatively, to end up with Type IIB theory one 
drops the dual coordinates $y^{\un{a}}$ and interprets what remains as the normal geometric 
coordinates. In the recent work \cite{Malek:2015hma} this procedure was used to obtain consistent 
truncations of Type IIA and IIB supergravities from the SL(5) extended geometry.  

With this in mind we return back to the decomposition \eqref{sl3decomp} and identify the 
$x_{\un{a}}$ with the geometric coordinates while the doublet $y^{\un{a}\hat{\a}}$ is identified 
with the doubled of winding modes for the fundamental string and the D1 brane. The latter are 
indeed dual with respect to an S-duality rotation. Finally the coordinate $z$ is understood as 
winding mode for the D3 brane in 3 dimensions. It is important to mention that the SL(2) symmetry 
is not broken.

To identify the fields of the SL(5) EFT let us look at the 7+3 decomposition of the (bosonic) 
fields of Type IIB supergravity
\begin{equation}
\label{IIB}
 \begin{aligned}
   & G_{\tt{M}\tt{N}} && \longrightarrow && g_{\m\n}, \, A_{\m \un{a}}, \, \f_{\un{ab}};\\
   & C_{\hat{\a}} && \longrightarrow && \f_{\hat{\a}};\\
   & B_{\tt{M}\tt{N} \hat{\a}} && \longrightarrow && B_{\m\n\,\hat{\a}}, \, A_{\m\, \un{a}\, 
\hat{\a}}, \, \f_{\un{ab}\, \hat{\a}};\\
   & C_{\tt{M}\tt{N}\tt{K}\tt{L}} && \longrightarrow &&  B_{\m\n\, \un{ab}}, \, A_{\m }, 
\, C_{\m\n\r\s}, \, C_{\m\n\r\, \un{a}}. 
 \end{aligned}
\end{equation}
Note that die to the self-duality of the 4-form gauge potential in 10 dimensions only the half of 
d.o.f. in the last line above survives. The branching rule imply the following decomposition of the 
EFT gauge fields
\begin{equation}
 \begin{aligned}
   && A_{\m}{}^{mn} & \to \{A_\m, \, A_{\m\un{a}}, \, A_{\m}{}^{\un{a}\hat{\a}}\},\\
   && B_{\m\n m} & \to \{B_{\m\n \un{a}}, \, B_{\m\n \hat{\a}}\}.
 \end{aligned}
\end{equation}
Considering only the fields $C_{\m\n\r\, \un{a}}$ and $A_{\m }$ of the last line in \eqref{IIB} as 
physical we identify them with the field $A_\m$ of EFT and the dual of $B_{\m\n \un{a}}$. Note 
that the underlined indices labeling the $\bf 3$ of SL(3) can be raised and lowered by the scalar 
matrix. With this in hand one directly identifies the remaining gauge fields.

The generalized metric $m_{mn}$ represented by the coset element can be decomposed as follows
\begin{equation}
 \begin{aligned}
   && m_{mn} & \to \{m_{\un{a}\un{b}},\, m_{\un{a}\hat{\a}},\, m_{\hat{\a}\hat{\b}}\}.
 \end{aligned}
\end{equation}
Here the fields $m_{\un{a}\hat{\a}}$ are directly identified with those coming from the 2-form in 
10 dimensions up to contraction with the alternating symbol $\e^{\un{a}\un{b}\un{c}}$. The element 
$m_{\un{a}\un{b}}$ of the coset GL(3)/SO(3) give the internal part $\f_{\un{a}\un{b}}$ of the 
10-dimensional metric, while the 2 fields $m_{\un{\a}\un{\b}}$ parameterizing the coset SL(2)/SO(2) 
match the axion-dilaton $C_{\hat{\a}}$. One may come to the same conclusions by analyzing the coset 
decomposition of the generalized vielbein, however we find the above analysis more transparent.

Hence, we conclude that the expected result of recovering the 11-dimensional supergravity and 
Type IIB supergravity by different solutions of the section condition holds for the SL(5) theory as 
for the other EFT's. The same procedure has been used in \cite{Blair:2013gqa} to explicitly obtain the 
Lagrangian for Type IIB and 11-dimensional supergravities from the internal sector of EFT, 
developed in \cite{Berman:2010is}  and \cite{Berman:2011jh}. One is still interested in doing the 
same for the full SL(5) EFT and for its supersymmetric extension.

\section{Discussion and outlook}

In this work the construction of SL(5) Exceptional Field Theory was presented, that fills the empty 
slot in the chain of EFT's for the groups $E_{8,7,6}$, SO(5,5) and SL(3)$\times$SL(2) already 
constructed. These correspond to the maximal supergravities in $D=3,4,5,6$ and $D=8$ respectively. 
Hence, the presented model adds the $D=7$ case and fulfills the chain. The U-duality groups for 
$D=9,10$ supergravity are too simple and the extended space can not be constructed. On the other end 
one meets the $E_9$ group to be expected as the U-duality group for $D=2$ maximal supergravity. This 
is infinitely dimensional, and hence the extended geometry in its known form ruins here as well.

For some applications, such as searches for solutions or classification of gaugings, models with 
SL(5) U-duality group seem to be more convenient as these provide less extended coordinates and carry more simple algebraic structure.

Although the initial construction of 
extended geometry has resulted from investigation of the toroidal backgrounds in supergravity, it is in 
general believed, that Scherk-Schwarz compactifications are able to catch non-toroidal 
and even non-geometric backgrounds. There was large progress in the direction of uplifting Type IIB 
solutions and solutions of 11-dimensional supergravity into EFT by choosing an appropriate Scherk-Schwarz 
reduction scheme \cite{Malek:2015hma,Baguet:2015sma,Baguet:2015xha}. However, there is still 
discussion in the literature, whether one should use other approaches to describe non-toroidal 
backgrounds. One of them is the so called WZW Double Field Theory, that attempts to construct a DFT 
on a group manifold \cite{Blumenhagen:2014gva,Bosque:2015jda,Blumenhagen:2015zma,Baguet:2015iou}. 
hence, in this context it would be interesting to expand the ideas of exceptional field theory to 
DFT${}_{WZW}$ and look for possible uplifts. 

The presented theory is essentially bosonic and one may be interested in extending it to include fermions in a supersymmetry invariant way. For the E${}_{6,7}$ EFT's this was done in \cite{Godazgar:2014nqa,Musaev:2014lna}. The interesting point here is that in contrast to the maximal gauged theories the bosonic Lagrangian is completely fixed already one the bosonic level with no need of supersymmetry. Hence, the fermionic sector should be constructed in such a way to fit nicely in the existing theory. With such theory in hands one may be able to investigate BPS solutions of the theory and geometry of Killing spinors.

In  \cite{Park:2014una} extended geometry for the group SL(N) was constructed, that may be interpreted as internal sector of a corresponding EFT. One may be interested in merging this work and the present results to end with an SL(N) ``exceptional'' field theory. The question is, to what extent one expect the known structures of tensor hierarchy to appear there, and is it possible to construct a Lagrangian.

\section*{Acknowledgements}

The author is grateful to Ilya Bakhmatov for valuable discussions and useful comments. The author would like to 
thank Nesin Mathematics Village (Izmir, Turkey) and its supporters for warm hospitality during 
completion of a  part of this research. This work on its final stage was supported by 
the Alexander von Humboldt Foundation.

\begin{appendix}

\section{Notations and conventions}\label{Notations}

All the notations for indices used in this paper are as follows
\begin{equation}
\begin{aligned}
& \hat{\tt{M}},\hat{\tt{N}},\ldots =0,\ldots 10, && \mbox{11-dimensional space-time indices};\\
& {\tt{M}},{\tt{N}},\ldots =0,\ldots 9, && \mbox{10-dimensional space-time indices};\\
& \m,\n,\r \ldots =0,\ldots 6, && \mbox{7-dimensional space-time indices};\\
& \ba,\bb,\bc \ldots =0,\ldots 6, && \mbox{7-dimensional space-time flat indices};\\
& a,b,c\ldots =1,\ldots 4, && \mbox{4-dimensional internal curved indices};\\
& \un{a}, \un{b},\un{c} \ldots =1,\ldots 4, && \mbox{4-dimensional internal curved Type IIB 
indices};\\
& \hat{\a} =1, 2, && \mbox{SL(2) Type IIB index};\\
& M,N,K \ldots =1,\ldots 10, && \mbox{indices of the $\bf 10$ of SL(5) labeling the extended 
space};\\
& m,n,k,l =1,\ldots 5, && \mbox{indices of the $\bf 5$ of SL(5)};\\
& \a,\b =1,\ldots 4, && \mbox{indices of the $\bf 4$ of USp(4)};
\end{aligned}
\end{equation}

The extended space of the SL(5) EFT is parametrized by the coordinates $\XX^M$ with the capital 
Latin indices labeling the representation $\bf 10$. However, it is often more convenient for 
explicit calculations to label the representation by an antisymmetric pair of indices in the 
fundamental $\XX^{mn}=-\XX^{mn}$. To prevent double counting one should either write sum 
with the condition $m<n$, or to write the 1/2 prefactor explicitly. To make the calculations more 
straightforward and more machine-friendly we choose the second way.

Hence, one observes the following rules to go from the capital Latin indices labeling the irrep 
$\bf 10$ to an antisymmetric pair of 
small Latin indices each labeling the irrep $\bf 5$
\begin{equation}
 \begin{aligned}
   T^M & \to T^{mn} && \mbox{any tensor};\\
   U^M V_M & \to \fr12 U^{mn}V_{mn};\\
   \d^M{}_N & \to  2 \d^{mn}_{kl} && \mbox{only for the Kronecker delta}.
 \end{aligned}
\end{equation}
The Kronecker delta symbol is required to be processed separately because $\d^{mn}_{mn}=10$ as 
well as $\d^M_M=10$, while one should introduce an extra $1/2$ factor when going from contraction 
of capital Latin indices to contraction of a pair of small indices.

\section{The algebra of SL(5)}
\label{algebra}

Generators of the SL(5) group in the fundamental representation and in the representation $\bf 
10$ are given by
\begin{equation}
\begin{aligned}
 (t^i{}_j)^m_n&=\d^m_j\d^i_n-\fr{1}{5}\d^m_n\d^i_j,\\
 (t^i{}_j)^{mn}{}_{kl}&=4(t^i{}_j)^{[m}_{[k}\d^{n]}_{l]}
 \end{aligned}
\end{equation}
These are traceless and satisfy the following commutation relations
\begin{equation}
 [t^m{}_n,t^k{}_l]=\d^m_lt^k{}_n-\d^k_nt^m{}_l.
\end{equation}
It is important to note, that when contracting generators in the $\bf 10$ representation one should 
use the capital Latin indices and the same is true for the projectors below. I.e. one writes
\begin{equation}
 (t^i{}_j t^k{}_l)^M{}_N=(t^i{}_j)^M{}_K (t^k{}_l)^K{}_N=\fr12(t^i{}_j)^M{}_{pq} (t^k{}_l)^{pq}{}_N.
\end{equation}
This results in a different coefficient in the definition of the generator in $\bf 10$ with respect 
to \cite{Berman:2012uy}, however we find such conventions more natural.

Now it is useful to write the explicit form of the projector on the $\bf 10$ representation of 
SL(5) that reads
\begin{equation}
 \PP^{M}{}_{N}{}^{K}{}_{L}=\fr13 (t^i{}_j)^{M}{}_{N}(t^i{}_j)^{K}{}_{L}.
\end{equation}
The identifying property of the projector then can be written as
\begin{equation} 
\PP^{M}{}_{N}{}^{K}{}_{L}\PP^{L}{}_{K}{}^{P}{}_{Q}=\fr14\PP^{M}{}_{N}{}^{kl}{}_{ij}\PP^{
ij}{}_{kl}{}^{P}{}_{Q} \overset{!}{=}\PP^{M}{}_{N}{}^{P}{}_{Q}.
\end{equation}
This fixes the overall prefactor in the projector and implies the correct identity
\begin{equation}
 \PP^{M}{}_{N}{}^{N}{}_{M}=\fr14\PP^{mn}{}_{kl}{}^{kl}{}_{mn}=24=\mbox{dim(adj)}.
\end{equation}

Let us now check explicitly the defining relation for the $Y$-tensor derived in \cite{Berman:2012vc}, that for 
the SL(5) group reads
\begin{equation}
 \e^{aMN}\e_{aKL}=Y^{MN}_{KL}=-3\PP^M{}_N{}^K{}_L+\fr15\d^M_N\d^K_L+\d^M_L\d^K_N,
\end{equation}
where $\e^{aMN}$ denotes the 5-dimensional alternating symbol $\e^{amnkl}$. Taking into account the 
above notations we rewrite the expression as
\begin{equation}
\e^{amnkl}\e_{apqrs}=-3\PP^{mn}{}_{pq}{}^{kl}{}_{rs}+\fr45\d^{mn}_{pq}\d^{kl}_{rs} 
+4\d^{mn}_{rs}\d^{kl}_{pq},
\end{equation}
note the prefactor $4$ of the Kronecker symbols. Substituting the expression for the projector in 
terms of the generators and writing them explicitly in terms of the Kronecker symbols we have 
for the RHS
\begin{equation}
 \begin{aligned} 
&\Big(-16(t^i{}_j)^{[m}{}_{[p}\d^{n]}_{q]}(t^j{}_i)^{[k}{}_{[r}\d^{l]}_{s]}\Big)+ 
\fr45\d^{mn}_{pq}\d^{kl}_{rs}+4\d^{mn}_{rs}\d^{kl}_{pq}\\
=&\Big(-8\d^{mn}_{q[r}\d^{kl}_{s]p}+8\d^{mn}_{q[r}\d^{kl}_{s]p}+\fr{16}{5}\d^{mn}_{pq}\d^{kl}_{rs}
\Big)+ \fr45\d^{mn}_{pq}\d^{kl}_{rs}+4\d^{mn}_{rs}\d^{kl}_{pq}\\
=&-8\d^{mn}_{q[r}\d^{kl}_{s]p}+8\d^{mn}_{q[r}\d^{kl}_{s]p}+4\d^{mn}_{pq}\d^{kl}_{rs}+4\d^{mn}_{rs}
\d^{kl}_{pq}=4!\d^{mnkl}_{pqrs}.
 \end{aligned}
\end{equation}
This is precisely what one has on the LHS of the identity, i.e. $\e^{amnkl}\e_{apqrs}=4!\d^{mnkl}_{pqrs}$.

\section{Useful identities}

The 4-form field strength $\mc{F}^{M,KL}$ belongs to the representation $\bf 10\otimes 
\bar{5}=5+45$ since the indices $KL$ by construction contain only the $\bf\bar{5}$. Moreover, the 
representation $\bf 45$ is not contained in the field strength. Hence, in the fundamental indices 
one may write $\mc{F}^{mn,[klrs]}$. And finally projecting out all the redundant representations 
one has 
\begin{equation}
\mc{F}^m=-6 \mc{F}^{mn,klrs}\e_{nklrs}.
\end{equation}

Consider now the following expression that is relevant for the Bianchi identity of the 3-form field 
strength $\mc{F}^{KL}$
\begin{equation}
\label{expr_bianchi}
\begin{aligned} 
 \e_{iKL}(\dt_N \mc{F}^{N,KL}-Y^{KL}_{PQ}\dt_N \mc{F}^{Q,PN})=\\
 \fr18(\dt_{mn}\mc{F}^{mn,klrs}\e_{iklrs}-6\e_{ipqrs}\dt_{mn}\mc{F}^{rs,pqmn}).
\end{aligned}
\end{equation}
Let us show, that this is proportional to $\dt_{ij}\mc{F}^j$. Indeed, considering 
antisymmetrization of the indices $\{niklrs\}$ in the first term, that is identically zero, and 
taking into account symmetries of the indices of $\mc{F}^{mn,klrs}$ we may write the following
\begin{equation}
 \begin{aligned}
\dt_{mn}\mc{F}^{mn,klrs}\e_{iklrs}=\dt_{mi}\mc{F}^{mn,klrs}\e_{nklrs}+ 
4\dt_{mk}\mc{F}^{mn,klrs}\e_{inlrs}.
 \end{aligned}
\end{equation}
Considering antisymmetrization of the indices $\{mnklrs\}$ in the second term above we rewrite the 
above expression as
\begin{equation}
\dt_{mn}\mc{F}^{mn,klrs}\e_{iklrs}=\dt_{mi}\mc{F}^{mn,klrs}\e_{nklrs}+ 
\fr12\dt_{mk}\mc{F}^{mk,nlrs}\e_{inlrs}+3\dt_{mk}\mc{F}^{nl,mkrs}\e_{inlrs}.
\end{equation}
Finally, substituting back into the expression \eqref{expr_bianchi} this gives the desired identity
\begin{equation}
\label{use_id1}
  \e_{iKL}(\dt_N \mc{F}^{N,KL}-Y^{KL}_{PQ}\dt_N \mc{F}^{Q,PN})=\fr{1}{24}\dt_{ij}\mc{F}^{j}.
\end{equation}
The Bianchi identity itself is then written as
\begin{equation}
 4\mc{D}_{[\m}\F_{\n\r\s] i}=6\e_{imnkl}\F_{[\m\n}^{mn}\F_{\r\s]}^{kl}+\dt_{mi}\F_{\m\n\r\s}{}^m.
\end{equation}

\section{Gauge invariance of the topological Lagrangian}
\label{gauge_top}

In this appendix explicit check of invariance of the topological Lagrangian under all gauge 
transformations is provided. Let us for convenience recall the variation of the topological 
Lagrangian
\begin{equation}
\label{var_top_App}
 \begin{aligned}
  \d \mc{L}_{top}= A\e^{\m\n\r\l\s\t\k}\bigg[\F_{\m\n\r\l}{}^{i}\dt_{ij} \D C_{\s\t\k}^{j}
+6\F_{\m\n}{}^{ij}\F_{\r\l\s i}\D B_{\t\k j}-2\F_{\m\n\r i}\F_{\l\s\t j}\d 
A_\k^{ij}\bigg].
 \end{aligned}
\end{equation}
Note, that the above expression is written completely in covariant terms, while this is not true 
for the topological Lagrangian itself. The only way to have a covariant form is to introduce 
a fictitious 8-dimensional space-time with a 7-dimensional and write a covariant expression, whose 
variation becomes a full derivative. Hence, one obtains an integration over the 7-dimensional 
boundary, that is formally identified with the usual space-time. 

Let us start first with generalized diffeomorphisms parametrized by $\L^M$, that give for 
\eqref{var_top_App} 
\begin{equation}
\begin{aligned}
\d_\L \mc{L}_{top}& \Rightarrow   \F_{\m\n\r\l}{}^m\dt_{mn}(\L^{nk}\F_{\s\t\k 
k})-12\e_{npqrs}\F_{\m\n}{}^{mn}\F_{\r\l\s m}\L^{pq}\F_{\t\k}{}^{rs}\\
&-2\F_{\m\n\r m}\F_{\l\s\t 
n}\DD_{\k}\L^{mn}.
\end{aligned}
\end{equation}
Here and everywhere in this section we omit the space-time alternating symbol $\e^{\m\n\r\l\s\t\k}$ 
to preserve space and for clarity of notations. Hence, the corresponding antisymmetrization of all 
the dummy space-time indices is always undermined. In addition, since one is actually dealing with 
the action rather than the Lagrangian, that involves integration over the space-time coordinates 
$x^\m$ as well as the extended coordinates $\XX^{mn}$, all full-derivative terms in $\DD_\m$ or 
$\dt_{mn}$ are dropped. 

Hence, performing integration by parts in the first term with respect to $\dt_{mn}$ and in the last 
term with respect to $\DD_\m$ and taking into account the hidden contraction with the alternating 
symbol we have
\begin{equation}
\begin{aligned}
& -\dt_{mn} \F_{\m\n\r\l}{}^m\L^{nk}\F_{\s\t\k k}-12\e_{npqrs}\F_{\m\n}{}^{mn}\F_{\r\l\s 
m}\L^{pq}\F_{\t\k}{}^{rs}+4(\DD_{\m}\F_{\n\r\l m})\F_{\s\t\k n}\L^{mn}\\
&=6\e_{mpqrs}\F_{\m\n}{}^{pq}\F_{\r\l}{}^{rs}\L^{mn}\F_{\s\t\k 
n}+12\e_{mpqrs}\F_{\m\n}{}^{mn}\F_{\r\l\s n}\L^{pq}\F_{\t\k}{}^{rs}\\
&=18\e_{mpqrs}\F_{\m\n}{}^{[pq}\F_{\r\l}{}^{rs}\L^{mn]}\F_{\s\t\k n}\equiv 0.
\end{aligned}
\end{equation}
Here we have used the Bianchi identity for $\F_{\m\n\r m}$ in the second line and organized a full 
antisymmetrization of six SL(5) indices $\{mnpqrs\}$ labeling the $\bf 5$ ensuring vanishing of 
the expression.

For gauge transformations parametrized by the 1-form parameter $\X_{\m k}$ we write
\begin{equation}
 \begin{aligned}
  \d_\X \mc{L}&\Rightarrow 3\F_{\m\n\r\l}{}^m\dt_{mn}(\F_{\s\t}{}^{nk}\X_{\k 
k})+12\F_{\m\n}{}^{mn}\F_{\r\l\s m}\DD_\t \X_{\k n}\\
&-\fr18 \F_{\m\n\r m}\F_{\l\s\t n}\e^{mnpqr}\dt_{pq}\X_{\k r}.
 \end{aligned}
\end{equation}
Performing integration by parts in all the terms and relabeling indices we obtain
\begin{equation}
 \begin{aligned}
   &-3\dt_{nm}\F_{\m\n\r\l}{}^n\F_{\s\t}{}^{mn}\X_{\k n}+12\F_{\s\t}{}^{mn}\DD_\m\F_{\n\r\l m} 
\X_{\k n}-12\DD_\t\F_{\m\n}{}^{mn}\F_{\r\l\s m} \X_{\k n}\\
&+\fr14 \dt_{pq}\F_{\m\n\r m}\F_{\l\s\t n}\e^{mnpqr}\X_{\k r}\\
&=18 
\e_{mpqrs}\F_{\s\t}{}^{mn}\F_{\m\n}{}^{pq}\F_{\r\l}{}^{rs}\X_{\k n}=18 
\e_{mpqrs}\F_{\s\t}{}^{[mn}\F_{\m\n}{}^{pq}\F_{\r\l}{}^{rs]}\X_{\k n}\equiv 0.
 \end{aligned}
\end{equation}
Here the second and the last terms in the first expression cancel dues to the Bianchi identity for 
the 2-form field strength $\F_{\m\n}{}^{mn}$, while the Bianchi identity for the 3-form field 
strength results in a single term. Following precisely the same arguments as above one observes 
the indices $\{ mnpqrs\}$ are fully antisymmetrized, and hence the term vanishes identically.

Finally, for the 3-form gauge transformations parametrized by $\Y_{\m\n}{}^m$ we have
\begin{equation}
 \begin{aligned}
   \d_\Y \mc{L}& \Rightarrow 3\F_{\m\n\r\l}\dt_{mn}\DD_\s \Y_{\t\k}{}^n-6\F_{\m\n}^{mn}\F_{\r\l\s 
m}\dt_{kn}\Y_{\t\k}{}^{k}\\
&=-3\DD_\s\F_{\m\n\r\l}\dt_{mn} \Y_{\t\k}{}^n-6\F_{\m\n}^{mn}\F_{\r\l\s 
m}\dt_{kn}\Y_{\t\k}{}^{k}\equiv 0.
 \end{aligned}
\end{equation}
Here in the first line we used the fact that $\dt_{mn}\DD_\s \Y_{\m\n}{}^n=\DD_\s\dt_{mn} 
\Y_{\m\n}{}^n$ and in the second line the Bianchi identity for the 4-form field strength. 

Let us now show that the used identity indeed holds, i.e. that one is allowed to swap derivatives 
in such expression. Effectively, this identity can be rewritten just as $\dt_{mn}\mc{L}_\L 
\Y^n-\mc{L}_\L\dt_{mn} \Y^n=0$ for some generalized tensor $\Y^m$ in the $\bf 5$ of SL(5) with 
generalized weight $\l[\Y]=3/5$. One notes, that this condition is nothing else but just a 
condition for $\dt_{mn}\Y^n$ to be a generalized tensor transforming under $\bf \bar{5}$. Hence, one 
indeed expects this to hold as precisely this term appears in the transformation $\D B_{\m\n m}$ 
and in the Bianchi identity for $\F_{\m\n\r m}$. Since all other terms in these expressions are 
generalized tensors the term $\dt_{mn}\Y^n$ should be a generalized tensor of weight 
$\l[\dt_{mn}\Y^n]=+2/5$.

However, let us check this explicitly and write first supposing $\dt_{mn}\Y^n$ is a generalized 
tensor of weight $+2/5$
\begin{equation}
 \begin{aligned}
   \dt_{mn}\mc{L}_\L 
\Y^n&=\fr12\dt_{mn}\L^{pq}\dt_{pq}\Y^n+\fr12\L^{pq}\dt_{mn}\dt_{pq}\Y^n-\fr14(t^n{}_r)^{kl}{}_{pq}
\dt_{mn}\dt_{kl}\L^{pq}\Y^r\\
&-\fr14(t^n{}_r)^{kl}{}_{pq}\dt_{kl}\L^{pq}\dt_{mn}\Y^r+\fr{3}{5\cdot 
2}\dt_{mn}\dt_{pq}\L^{pq}\Y^n+\fr{3}{5\cdot 2}\dt_{pq}\L^{pq}\dt_{mn}\Y^n;\\
\mc{L}_\L 
(\dt_{mn}\Y^n)&=\fr12\L^{pq}\dt_{pq}\dt_{mn}\Y^n+\fr14(t^r{}_{m})^{kl}{}_{pq}\dt_{kl}\L^{pq} 
\dt_{rn}\Y^n+\fr{2}{5\cdot 2}\dt_{pq}\L^{pq}\dt_{mn}\Y^n.
 \end{aligned}
\end{equation}
We now show that these expressions are equivalent up to terms vanishing under the section 
condition. Taking difference of these expressions one notes, that there are terms of the form 
$\dt\L\dt\Y$ and $\dt\dt\L \Y$ that should vanish separately. Indeed, we have for the first type
\begin{equation}
 \begin{aligned}
&\Rightarrow \fr12\dt_{mn}\L^{pq}\dt_{pq}\Y^n-\fr14(t^n{}_r)^{kl}{}_{pq}\dt_{kl}\L^{pq}\dt_{mn}
\Y^r-\fr14(t^r{}_{m})^{kl}{}_{pq}\dt_{kl}\L^{pq} 
\dt_{rn}\Y^n+\fr { 1 } { 10}\dt_{pq}\L^{pq}\dt_{mn}\Y^n\\
&=\fr12\dt_{mn}\L^{pq}\dt_{pq}\Y^n-\dt_{nq}\L^{pq}\dt_{mp}\Y^n-\dt_{mq}\L^{pq}\dt_{pn}\Y^n+ 
\fr12\dt_{pq}\L^{pq}\dt_{mn}\Y^n\\
&=3\dt_{[mn}\L^{pq}\dt_{pq]}\Y^n\equiv 0,
 \end{aligned}
\end{equation}
where in the second line the explicit form of the SL(5) generators and in the last line 
the section condition were used.

The similar calculation can be performed for the terms of the second type and one gets the following
\begin{equation}
  \begin{aligned}
&\Rightarrow-\fr14(t^n{}_r)^{kl}{}_{pq}\dt_{mn}\dt_{kl}\L^{pq}\Y^r+\fr{3}{10}\dt_{mr}\dt_{pq}\L^{pq}
\Y^r\\
&=-\dt_{mp}\dt_{rq}\L^{pq}\Y^r+\fr12\dt_{mr}\dt_{pq}\L^{pq}\Y^r=-3\dt_{[mp}\dt_{rq]}\L^{pq}
\Y^r\equiv 0.
  \end{aligned}
\end{equation}

Hence, this concludes the explicit proof of invariance of the topological Lagrangian. This 
invariance fixes all internal coefficients, while leaving the overall prefactor arbitrary. The 
latter will be fixed by invariance under external 1+6--dimensional diffeomorphisms.

\end{appendix}

\bibliographystyle{utphys}
\bibliography{bib}
\end{document}

%% file: Defs.tex
\def\a{\alpha} \def\b{\beta} \def\g{\gamma} \def\d{\delta} \def\e{\epsilon}
\def\ve{\varepsilon}   
  \def\k{\kappa} \def\l{\lambda} \def\m{\mu}
\def\n{\nu} \def\x{\xi}   \def\r{\rho}
 \def\s{\sigma} \def\t{\tau}  \def\f{\varphi}
\def\ff{\phi} \def\c{\chi} \def\y{\psi} \def\w{\omega}

\def\D{\Delta} 
   \def\Q{\Theta}
   \def\L{\Lambda} 
 \def\X{\Xi}  
 \def\S{\Sigma}  
\def\F{\Phi}  \def\Y{\Psi} \def\W{\Omega}

\def\ba{\bar{a}}\def\bb{{\bar{b}}}\def\bc{\bar{c}}
\def\bp{\bar{p}}\def\bq{\bar{q}}

\def\fr{\frac}  \def\dt{\partial}

\def\ph{\phantom}
\def\mc{\mathcal}

\def\XX{\mathbb{X}}

\newcommand\bqa {\begin{eqnarray}}
\newcommand\eqa {\end{eqnarray}}

\newcommand{\bear}{\begin{array}}
\newcommand{\enar}{\end{array}}

\newcommand{\un}[1]{\underline{#1}}

\def\beq{\begin{equation}}
\def\eeq{\end{equation}}
\def\bea{\begin{eqnarray}}
\def\eea{\end{eqnarray}}

\def\F{{\mathcal{F}}}

\def\PP{\mathbb{P}}
\def\LL{\mathcal{L}}
\def\M{{\mathcal{M}}}
\def\DD{{\mathcal{D}}}